\pdfoutput=1
\documentclass[journal]{IEEEtran}
\usepackage{amsmath,xcolor}
\usepackage{graphicx}
\usepackage{cite}
\usepackage{epstopdf}
\usepackage{url}
\usepackage{amsthm}
\usepackage{amssymb}
\usepackage{amsfonts}
\usepackage{subfigure}
\usepackage{algorithm}

\usepackage{algorithmicx}
\usepackage[noend]{algpseudocode}
\title{\LARGE \bf Fault Tolerant Control of Automotive Air Conditioning Systems using a GIMC Structure}
\author{Xu Zhang

\thanks{\quad Xu Zhang is with Department of Control Science and Engineering, Harbin Institute of Technology, Harbin, 150080, China {\tt\small
{zhangxu\_hit@126.com}}

}
}
\IEEEoverridecommandlockouts

\graphicspath{{figures/}}

\begin{document}
\maketitle

\begin{abstract}
Although model-based fault tolerant control (FTC) has become prevalent in various engineering fields, its application to air-conditioning systems is limited due to the lack of control-oriented models to characterize the phase change of refrigerant in the vapor compression cycle. The emergence of moving boundary method (MBM) illuminates a promising way for FTC design. In this paper, we exploit a control-oriented nonlinear model comparable to MBM to design an FTC framework with a generalized internal model control (GIMC) approach. A fault detector and isolator (FDI) is developed to identify potential actuator and sensor faults. A fault compensator is employed to compensate these faults if detected.  Comprehensive simulations are carried out to evaluate the developed FTC framework with promising results. Plant variations are explicitly considered to enhance the gain-scheduled FTC developments.
\end{abstract}
\begin{IEEEkeywords}
Air Conditioning System, Fault Tolerant Control, Generalized Internal Model Control, Moving Boundary Method, Robust Control
\end{IEEEkeywords}

\section{Introduction}
Fault tolerant control (FTC) has attracted much attention from control engineers since it can accommodate sensor and actuator faults to preserve desired close-loop performance \cite{50_chen2012, 50_isermann1997, 50_isermann2006, 50_blanke2006, 10_zhang2008}. FTC schemes mainly fall into two categories: passive and active. Passive approaches aim to retain stability and performance against all faults by a single controller. One the other hand, an active approach typically consists of a fault diagnosis stage followed by a controller reconfiguration. Although active approaches require more computational power during implementation, they typically yield less conservative results and better closed-loop performance when faults occur.

Recently, great research attention has been given to the automotive industry to achieve improved handling, comfort and safety, as well as route planning, road condition assessment, and pothole detection capability, fuel efficiency prediction, etc. \cite{J2,J3,6214702,J8,C7,C1,C2,1219456}. In control-related areas, various model-based control and estimation approaches have been widely developed \cite{C3,J7,7506101,C9,7533442,J9,C4}. However, model-based FTC design for automotive Air Conditioning (A/C) system has rarely been touched. The main reason of this stagnation is the lack of a control-oriented model that can balance model accuracy and computational complexity to characterize the thermo-fluid dynamics of the refrigerant \cite{10_katipamula2005p1}. During heat removal/release at the heat exchanger,  refrigerant experiences a liquid/vapor phase change. This phase change is a complicated process in which mass and energy balance is difficult to characterize. Hence, previous studies were based on static or empirical models where the performance after fault occurrences was unsatisfactory in transient \cite{10_keir2006}.

Although steady-state analysis is the main stream in FTC, studies have recently emerged on dynamic fault diagnosis \cite{10_janecke2011}, in which transient data and models are used to identify faulty system behaviors. For example, a lumped parameter method has been applied to a vapor compression cycle with a fixed orifice device to develop an observer-based scheme \cite{10_wagner1992}. Black box models obtained through data-driven techniques, such as ARX and ARMAX, were used to generate structured residuals to predict faults in an air handling unit \cite{10_lee1996}. In \cite{10_keir2006}, a comprehensive model was used to represent the vapor compression cycle as opposed to an air handing unit and a linearized model was used to explore the sensitivity of each output to a variety of faults. However, no practical fault diagnosis algorithm was implemented and tested.

In summary, model-based FTC design for automotive A/C systems is far from satisfactory on three aspects. Firstly, the vapor compression cycle, a critical subsystem in the A/C system for energy conversion, has not been comprehensively studied; previous work focused on the air handling unit. Secondly, lumped parameter modeling and data-driven modeling approaches are not physics-based, rendering little understanding of the relationships between possible faults and corresponding symptoms. Thirdly, a gap exists between fault diagnosis and control design, from the fact that a seamlessly integrated design approach combining both fault diagnosis and control action has not been investigated yet. In this paper, we aim to provide a benchmark for dynamic fault diagnosis and control of vapor compression cycle in a unified framework, by merging the latest advances in both practitioner and theory developments.

A promising approach to develop control-oriented models for heat exchangers was proposed in \cite{01_he1997, 01_Asada1998,02_Li2010} by exploiting the Moving Boundary Method (MBM). The refrigerant is lumped based on its phase status, i.e., pure vapor, pure liquid or mixed vapor and liquid. By exploiting mean void fraction and volumetric ratio of vapor over liquid, a set of differential equations describing the mass, momentum, and energy balances of the phase change process was developed and solved. This model is advantageous in modeling the transient behavior, and it is more precise than existing modeling techniques. The use of a first-principle A/C model can significantly reduce the time required to develop and implement FTC algorithms.

As we mentioned, active FTC schemes rely on fault detection to introduce fault-feedback compensation or control reconfiguration based on identified fault information. For instance, an integrated control and fault detection design, based on a four parameter controller, was proposed in \cite{10_nett1988, 10_niemann1997, 10_stoustrup1997,10_marcos2005}, which incorporates additional two Degrees of Freedom (DoFs) for the purpose of fault diagnosis. An alternative implementation of the integrated control, referred to as Generalized Internal Model Control (GIMC), has been introduced in \cite{10_zhou2001,10_campos2003,10_campos2005, 10_campos2008}, which is able to overcome conflicts between performance and robustness in traditional feedback frameworks. With GIMC, a high performance controller is active under normal conditions and a robust controller will be activated when sensor/actuator faults or external disturbances are identified.

In this paper, MBM A/C modeling and FTC using the GIMC structure are integrated in a unified framework. The contributions of this paper include the following. First, a first-principle vapor compression cycle model is exploited to model the complex thermo-fluid process. Furthermore, a fault tolerant control scheme is developed by using the GIMC method. Finally, a gain-scheduling compensator for fault accommodation is developed and simulations are presented to demonstrate the efficacy of the proposed framework.

The rest of the paper is organized as follows. The fundamental theory of FTC with GIMC structure is detailed in Section II. Mathematical modeling of the A/C plant and gain-scheduled $H_{\infty}$ controller design are introduced in Section III. Fault detection and isolation algorithm for sensor and actuator faults is developed in Section IV and fault tolerant controller is designed in Section V with a preliminary study on the gain-scheduled GIMC structure. Conclusion remarks are made in Section VI.

\section{Fault Tolerant Control (FTC) Scheme}
In this section, a general FTC scheme for automotive A/C systems is introduced with individual modules detailed. Integration of control action and reconfiguration mechanism is realized through a method called the GIMC structure that can actively reconfigure the controller once faults occur.
\subsection{Controller Reconfiguration}
\begin{figure*}[!htb]
\centering
\includegraphics[width=0.8\textwidth]{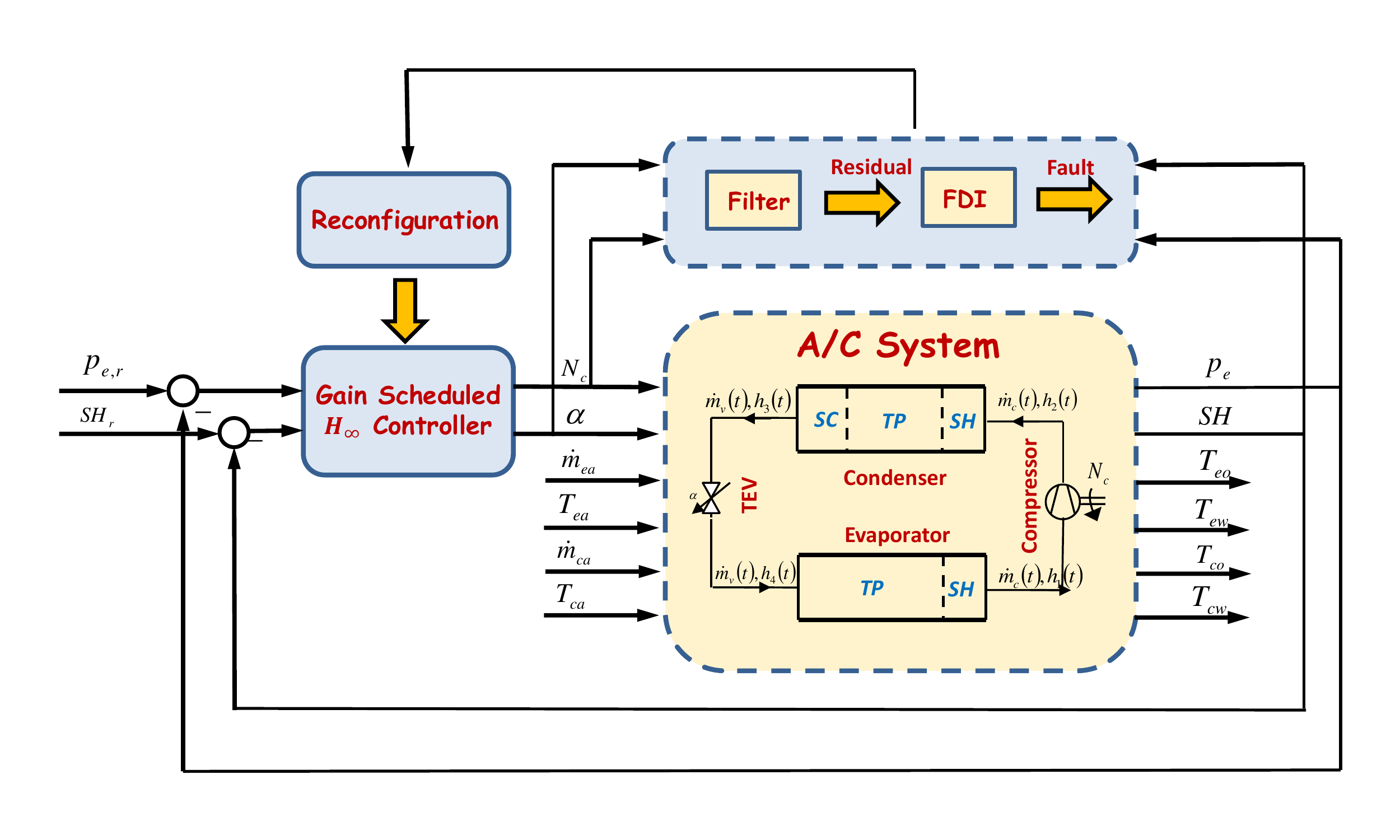}
\caption{Interconnections of Plant, Controller, FDI and Reconfiguration.}
\label{fig:ClosedLoop}
\end{figure*}

The FTC scheme of an automotive A/C system is illustrated in Figure \ref{fig:ClosedLoop}. The scheme consists of four modules:  A/C plant, gain-scheduled controller, Fault Detection and Isolation (FDI), and reconfiguration mechanism.

As illustrated in Figure \ref{fig:ClosedLoop}, a basic automotive A/C system is composed of four primary components, evaporator, compressor, condenser and expansion valve. The vapor compression cycle removes heat from the air flowing into the cabin through the evaporator, as the refrigerant evaporates from two-phase (TP) status into superheated (SH) status, and rejects heat to the air flowing through the condenser, as the refrigerant condenses from superheated (SH) status into sub-cooled (SC) status through two-phase (TP) status. The enthalpy, mass flow rate and pressure, are exchanged via the four components. Basically, the two heat exchangers set the pressures of the system, while the compressor and expansion valve determine the mass flow rates at the inlet and outlet of the evaporator and condenser.

From a system-level perspective, the interfaces of the A/C plant to the rest of the scheme are:
\begin{enumerate}
  \item The controller sends commands to the A/C plant through the two controllable inputs, the compressor speed $N_c$ and the valve position $\alpha$,
  \item Two measurements are available for the controller and the FDI module, the evaporator pressure $p_e$ and the superheat temperature $SH$ .
\end{enumerate}
The air mass flow rates and temperatures on the evaporator side $\dot{m}_{ea}, T_{eo}$ and condenser side $\dot{m}_{ca}, T_{co}$, are measurable but noncontrollable disturbances. The air temperatures leaving the heat exchangers $T_{eo},T_{co}$ and tube wall temperatures $T_{ew}, T_{cs}$ are calculated using a control-oriented A/C model.

A controller is designed to track prescribed trajectories of two output variables, namely the evaporator pressure $p_e$, and the superheat temperature $SH$ \cite{10_zhang2014}. The reference values for the tracked variables are labeled as $p_{e,r}$ and $SH_r$, respectively, which are generated by higher level optimization algorithms developed for improving energy conversion efficiency. Meanwhile, the controller is expected to reject disturbances caused by the variation of air mass flow rate at the evaporator, $\dot m_{ea}$, which is manually set by the driver or intelligently regulated by the cabin control unit. At different cooling loads, the coupling between inputs and outputs change significantly. Hence, the controller is supposed to be scheduled according to the system operating condition.

An FTC scheme is targeted to achieve stability and performance, not only when all modules function normally, but also in cases when there are faults in sensors, actuators, or other system components \cite{10_zhang2008}. The faults of interest are one actuator fault (compressor speed) and one sensor fault (pressure reading). Both are assumed to enter the A/C system additively, meaning that 1) the actual compressor speed $N_{cmp,act}$ is the sum of the commanded compressor speed $N_{cmp}$ sent by the controller and a faulty compressor speed $f_N$; 2) the pressure measurement $p_{e,msr}$ available to both the controller and FDI module is the actual pressure in the system $p_e$ plus a faulty reading $f_p$.

If the FDI module and reconfiguration mechanism are absent, the FTC scheme is considered passive as the controller is pre-determined in the design phase. Since it aims to be robust against a class of presumed faults, the fault-tolerant capabilities are limited. In contrast, an FTC scheme with both the FDI module and reconfiguration mechanism is considered active, because it reacts to faults actively by reconfiguring control actions so that the stability and performance of the closed-loop system can be preserved. For a successful control reconfiguration, a real-time FDI module is required to provide precise information of the faulty components in the system. In the FDI module, both commanded control inputs and measurement outputs from the A/C plant are synthesized to generate residuals which are signals essential for fault detection, isolation and estimation.
\subsection{GIMC Structure}
\begin{figure}[!htb]
\centering
\includegraphics[width=0.49\textwidth]{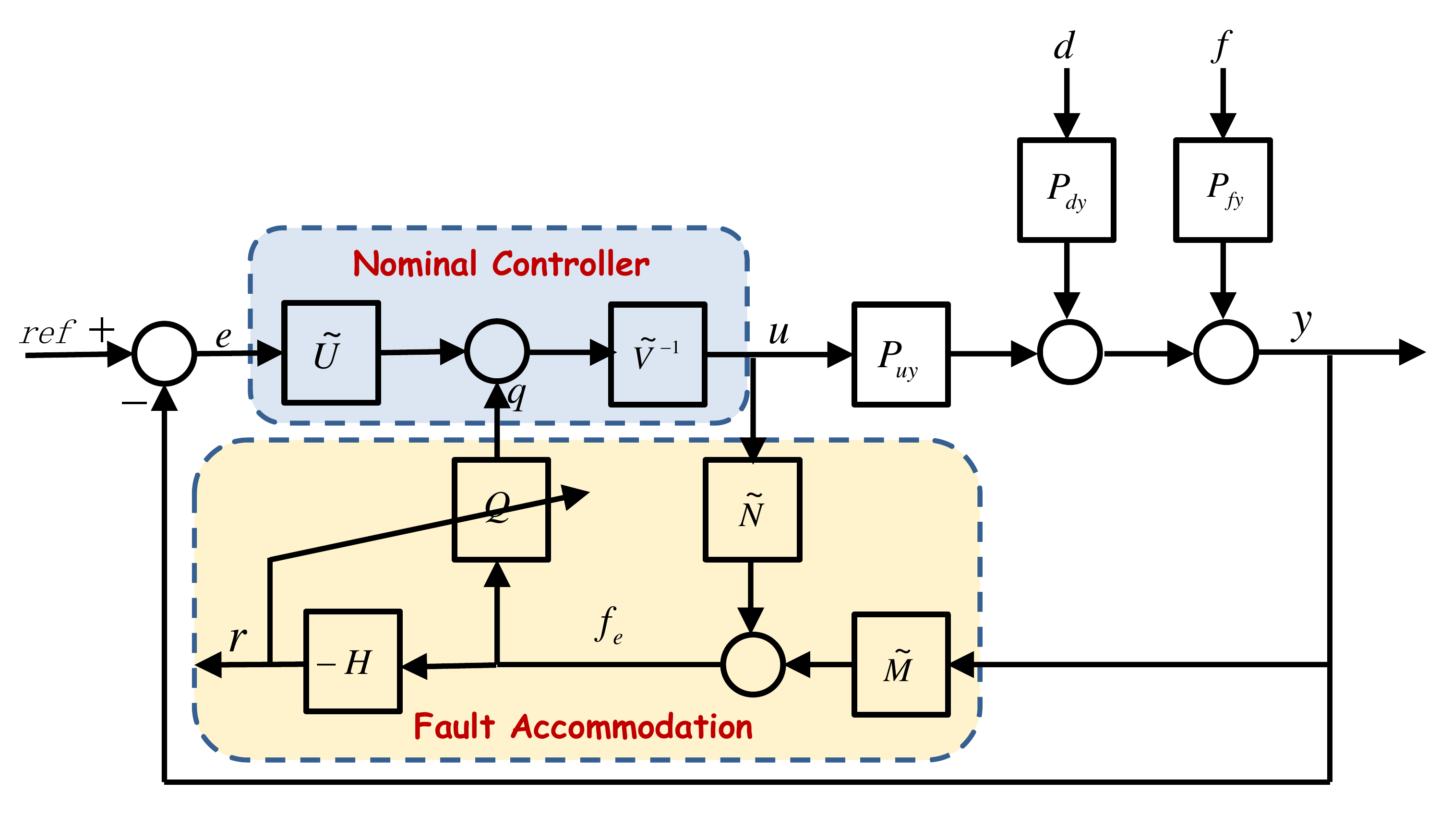}
\caption{Generalized Internal Model Control Structure (Adapted from \cite{10_zhou2001}).}
\label{fig:GIMC_Scheme}
\end{figure}

The general FTC scheme in Figure \ref{fig:ClosedLoop} relies on an FDI algorithm, followed by a fault accommodation into the nominal controller. The FDI module is expected to detect and isolate the occurrence of a fault in the closed-loop system, and provide an appropriate compensation signal to the controller in order to maintain the closed-loop performance. The general FTC scheme is realized through an integrated FDI design and reconfiguration mechanism referred to as the GIMC structure \cite{10_zhou2001}, as shown in Figure \ref{fig:GIMC_Scheme}, in which design methods for nominal conditions \cite{10_campos2008} are summarized below for reader's convenience.

Consider a linear system $P(s)$ affected by disturbances $d \in \mathbb{R}^r$ and possible faults $f \in \mathbb{R}^f$ described by
\begin{equation}
    \left \{
      \begin{array}{l}
        \dot{x} = Ax + Bu + F_1 f + E_1 d, \\
        y = Cx + Du + F_2 f + E_2 d, \\
      \end{array}
    \right.
\end{equation}
where $x \in \mathbb{R}^n$ represents the vector of states, $u \in \mathbb{R}^m$ is the vector of inputs, and $y \in \mathbb{R}^p$ represents the vector of outputs. The nominal system is considered to be controllable and observable. The system response $y$ can be analyzed in a transfer matrix form as follows:
\begin{equation}
    y = P_{uy} u(s) + P_{fy} f(s) + P_{dy} d(s).
\end{equation}
A left coprime factorization for each transfer matrix can be derived as:
\begin{equation}
    P_{uy} = \tilde{M}^{-1} \tilde {N},\, P_{dy} = \tilde{M}^{-1} \tilde {N}_d,\, P_{fy} = \tilde{M}^{-1} \tilde {N}_f,
\end{equation}
where $\tilde{M}, \tilde {N}, \tilde {N}_d, \tilde {N}_f \in RH_{\infty}$.

Now suppose a nominal controller $K$ that stabilizes the nominal plant $P_{uy}$, and provides a desired closed-loop performance in terms of robustness, transient, and steady state responses. The controller can be represented by a left coprime factorization,
\begin{equation}
    K = \tilde{V}^{-1} \tilde{U}
\end{equation}

The accommodation scheme adopted in this paper is motivated by a new implementation of the Youla parametrization referred to as GIMC \cite{10_zhou2001}. In this configuration, it allows the system to perform FDI and fault accommodation in the unified structure, where these two processes can be carried out by selecting two design parameters $Q, H \in RH_{\infty}$. Consequently, the residual $r$ is generated by selecting the detection/isolation filter $H$ and the accommodation signal $q$ is generated by the compensator $Q$, using
the filtered signal $f_e$ with the following criteria \cite{10_campos2008}.
\begin{itemize}
  \item $H(s)$: the fault detection/isolation filter must diminish the effect of the disturbances or uncertainty into the residual signal, and maximize the effect of the faults.
  \item $Q(s)$: the robustification controller must provide robustness into the closed-loop system in order to maintain acceptable performance against faults.
\end{itemize}

The GIMC structure functions as follows: $r=0$ if there is no model uncertainties, external disturbances or faults and then the control system will be solely controlled by the high performance controller $K_0 = \tilde{V}^{-1} \tilde{U}$. On the other hand, the robustification controller $Q$ will only be active when $r \neq 0$, i.e., there are either model uncertainties or external disturbances or sensor/actuator faults. The advantage of the GIMC structure is that if there is no uncertainty, the controller will perform as well as a nominal controller does; if uncertainty exists, the controller implementation should in principle perform no worse than the standard robust controller implementation does as to robustness and performance.

\section{A/C Plant}
Realization of the general FTC scheme using the GIMC structure needs left coprime factorization of the A/C plant model in order to design the detection/isolation filter $H$ and the compensator $Q$. Here, the A/C plant described using MBM language is utilized to generate a control-oriented model that not only provides the $\tilde{M}$ and $\tilde{N}$ matrices for design purpose, but also serves as a nonlinear simulator for algorithm validation.

In the MBM modeling framework, the compressor and the valve are modeled as static components. The dynamics related to the heat and mass transfer inside the heat exchangers are described using the MBM method \cite{01_he1997, 02_Li2010}, where Reynolds transport theorem describing the mass and energy conservation for transient one-dimensional flow is applied to each phase region of the condenser and evaporator with boundary conditions and refrigerant properties specified. After derivations detailed in \cite{50_zhang2014JDSMC} and not included here for brevity, the final mathematical equations describing system dynamics are in the descriptor form,
\begin{equation}
\label{E:Descriptor_Form2}
\begin{aligned}
  Z(x,f_a) \frac{dx}{dt} &= f(x,f_a,u,v,f_N), \\
  y&=g(x,f_a,f_p),
\end{aligned}
\end{equation}
where $f_N$ and $f_p$ are, respectively, the compressor and the pressure fault as aforementioned. The input vector $u$ includes the compressor rotation speed and expansion valve opening percentage, i.e., $u = \begin{bmatrix} N_{c} & \alpha \end{bmatrix} ^T$. The boundary conditions are the variables describing the air side of the heat exchangers, and could be treated as unknown disturbances, $ v =  \begin{bmatrix}\dot{m}_{ea}  & T_{ea,in}  \end{bmatrix} ^T $. The state vector $x_e$, describing the evaporator status, includes 5 variables, $ x_e= \begin{bmatrix} \zeta_{e1} & p_e & h_{e2} & T_{e1w} & T_{e2w}  \end{bmatrix} ^T $.  Finally, the output vector $y$ includes the evaporator pressure and superheat temperature, $y =\begin{bmatrix} p_e & SH \end{bmatrix}^T$, which are algebraic functions of states, $g(x)$. The $Z$ matrix and $f$ vector are complex expressions of refrigerant properties, heat transfer coefficients and geometric parameters \cite{50_zhang2014JDSMC}.

\subsection{Mathematical Model}
The compressor and expansion valve are the two main actuators regulating the pressure difference and enthalpy distribution in the A/C loop. In the compressor, the mass flow rate $\dot m_c$ and outlet enthalpy $h_2$ are, respectively,defined as:
\begin{equation}
\label{E:CMP_E}
\begin{aligned}
   \dot m_c &= \eta_v V_d \rho_1 \omega_c ,  \\
   h_2 &= \frac {h_{2s} - h_1}{ \eta_s} + h_1,
   \end{aligned}
\end{equation}
where $V_d$ is the compressor displacement; $\rho_1, h_1$ are the refrigerant density and enthalpy at the compressor inlet, respectively; $\omega_c$ is the compressor speed and $h_{2s}-h_1$ is the isentropic enthalpy difference. The first control input is the compressor rotation speed $N_c$ in the unit of $rpm$.

The mass flow rate through the expansion valve is modeled by the orifice flow equation, approximated by assuming constant fluid density:
\begin{equation}
\label{E:TEVM}
  \dot m_v = C_{d,v} A_v \sqrt{2\rho_3\left(p_{3} - p_{4}\right)},
\end{equation}
where $A_v$ is the valve curtain area and $C_v$ is the discharge coefficient. The outlet enthalpy is typically found by assuming an ideal throttling process, hence $h_4 = h_3$. The second control input is the valve position $\alpha$ in percentage, determining the effective flow area of the valve.

The mass and energy balance equations for the two-phase region of the evaporator are given directly in Equations \ref{E:evap_3} and \ref{E:evap_4}, respectively,
\begin{figure*}[!hbt]
\small
\begin{equation}
\label{E:evap_3}
\begin{aligned}
  &\left(\frac{\rho_{e,TP}-\rho_g}{\rho_{e,TP}}\right)\frac{d\zeta_1}{dt} + \frac{1}{\rho_{e,TP}}\frac{\partial \rho_{e,TP}}{\partial p_e}\frac{dp_e}{dt}\cdot\zeta_1 + \frac{1}{\rho_{e,TP}}\frac{\partial \rho_{e,TP}}{\partial \bar\gamma_e}\frac{d\bar\gamma_e}{dt}\cdot\zeta_1\\
  &= \frac{\dot m_v}{\rho_{e,TP}V_e} - \frac{\dot m_{12}}{\rho_{e,TP}V_e}
  \frac{\rho_g \left(h_{e,TP}-h_g\right)}{\rho_{e,TP}}\frac{d\zeta_1}{dt} + \left(\frac{\partial h_{e,TP}}{\partial p_e}-\frac{1}{\rho_{e,TP}}\right)\frac{dp_e}{dt}\cdot\zeta_1 + \frac{\partial h_{e,TP}}{\partial \bar\gamma_e}\frac{d\bar\gamma_e}{dt}\cdot\zeta_1  \\
  & = \frac{\dot m_v}{\rho_{e,TP}V_e} \left(h_4-h_{e,TP}\right) - \frac{\dot m_{12}}{\rho_{e,TP}V_e} \left(h_g-h_{e,TP}\right) +\frac{\dot{Q}_{TP}}{\rho_{e,TP}V_e},
\end{aligned}
\end{equation}
\normalsize
\end{figure*}
\begin{figure*}[!htb]
\small
\begin{equation}
\label{E:evap_4}
\begin{aligned}
  &-\left(\frac{\rho_{e,SH}-\rho_g}{\rho_{e,SH}}\right)\frac{d\zeta_1}{dt} + \frac{1}{\rho_{e,SH}}\frac{\partial \rho_{e,SH}}{\partial p_e}\frac{dp_e}{dt}\cdot\left(1-\zeta_1\right) + \frac{1}{\rho_{e,SH}}\frac{\partial \rho_{e,SH}}{\partial h_{e,SH}}\frac{dh_{e,SH}}{dt}\cdot\left(1-\zeta_1\right) \\
  & = \frac{\dot m_{12}}{\rho_{e,SH}V_e} - \frac{\dot m_{c}}{\rho_{e,SH}V_e}-\frac{\rho_g \left(h_{g}-h_{e,SH}\right)}{\rho_{e,SH}}\frac{d\zeta_1}{dt} + \frac{1}{\rho_{e,TP}}\frac{dp_e}{dt}\cdot\left(1-\zeta_1\right) - \frac{dh_{e,SH}}{dt}\cdot\left(1-\zeta_1\right)  \\
  & = \frac{\dot m_{12}}{\rho_{e,SH}V_e} \left(h_g-h_{e,SH}\right) - \frac{\dot m_{c}}{\rho_{e,SH}V_e} \left(h_1-h_{e,SH}\right) +\frac{\dot{Q}_{SH}}{\rho_{e,SH}V_e},
\end{aligned}
\end{equation}
\normalsize
\end{figure*}
where $p_e$ is the evaporator pressure; $\zeta_1$ is the two-phase region normalized tube length; $h_{e,SH}$ is the enthalpy of the refrigerant at the tube exit; $\dot{m}$ is the mass flow rate; $\dot{Q}$ is the heat transfer rate; $\rho$ denotes the density. In Equations \ref{E:evap_3} and \ref{E:evap_4}, the left hands represent the variation of independent states of the refrigerant, and the right hands the exchanger of mass and energy at the inlet and outlet of individual phase region, as well as the heat transfer along the wall of corresponding region. The terms multiplying the state variations depend on the refrigerant inherent thermodynamic properties, hence are state-dependent. The mass and energy balances for the sub-cooled, two-phase and superheated region of the condenser are not shown here for brevity.

\subsection{Coprime factorization of Plant Model}
\begin{figure}[!htb]
    \centering
      \subfigure[Vehicle Speed Profile]
      {\includegraphics[width=0.45\textwidth]{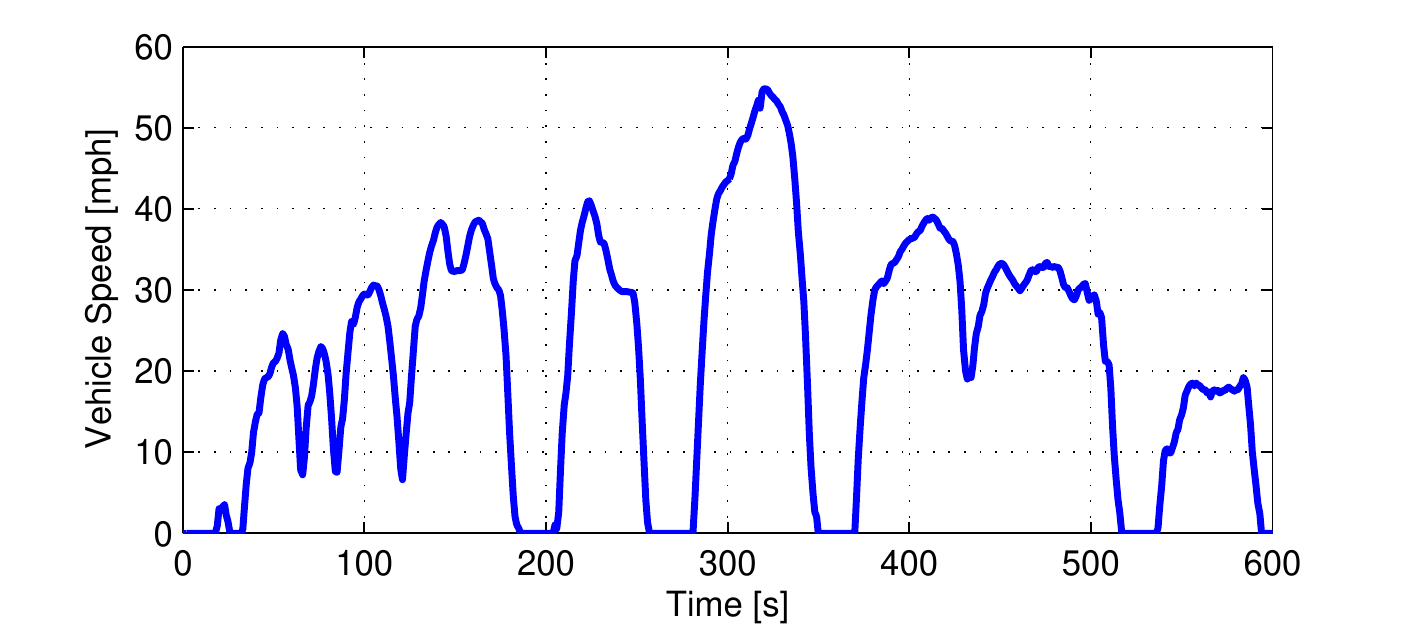}
      \label{fig:Veh_SC03}}
      \subfigure[Condenser Pressure]
      {\includegraphics[width=0.45\textwidth,keepaspectratio=true]{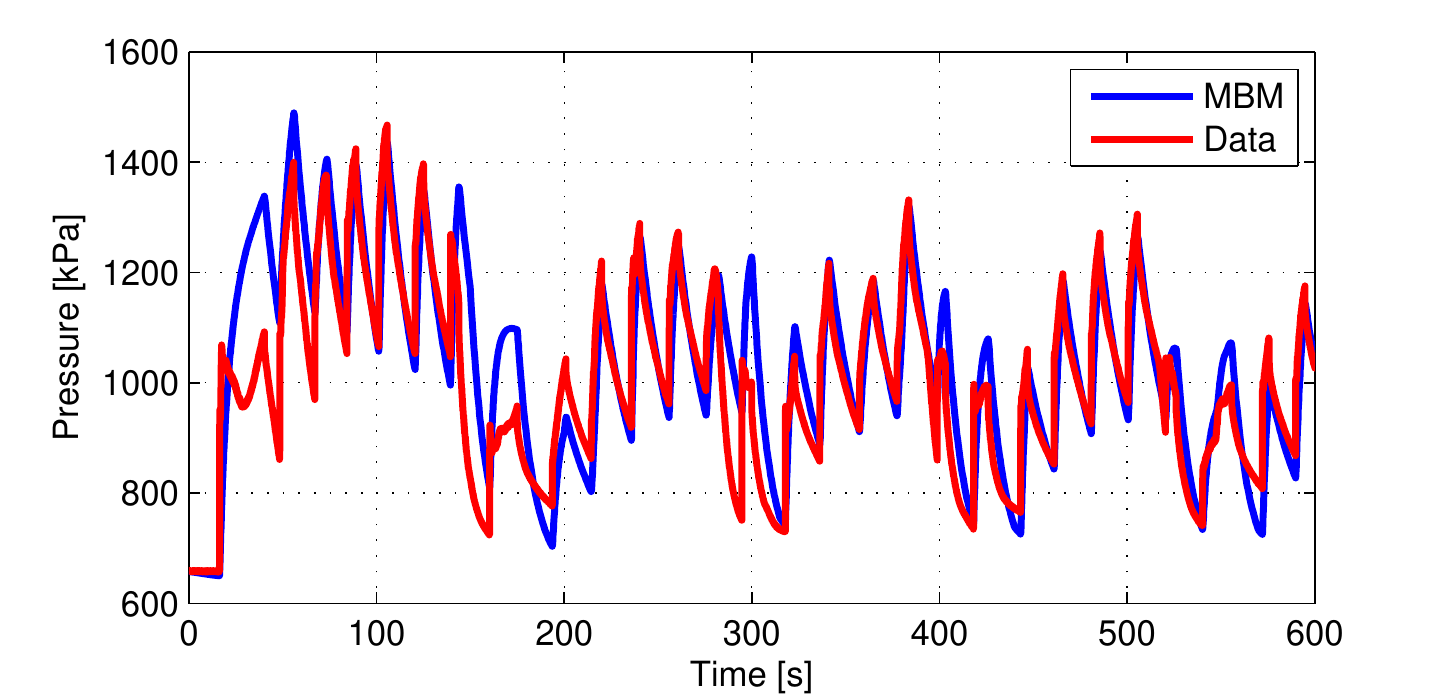}
      \label{fig: Pc_SC03}}
      \subfigure[Evaporator Pressure]
      {\includegraphics[width=0.45\textwidth,keepaspectratio=true]{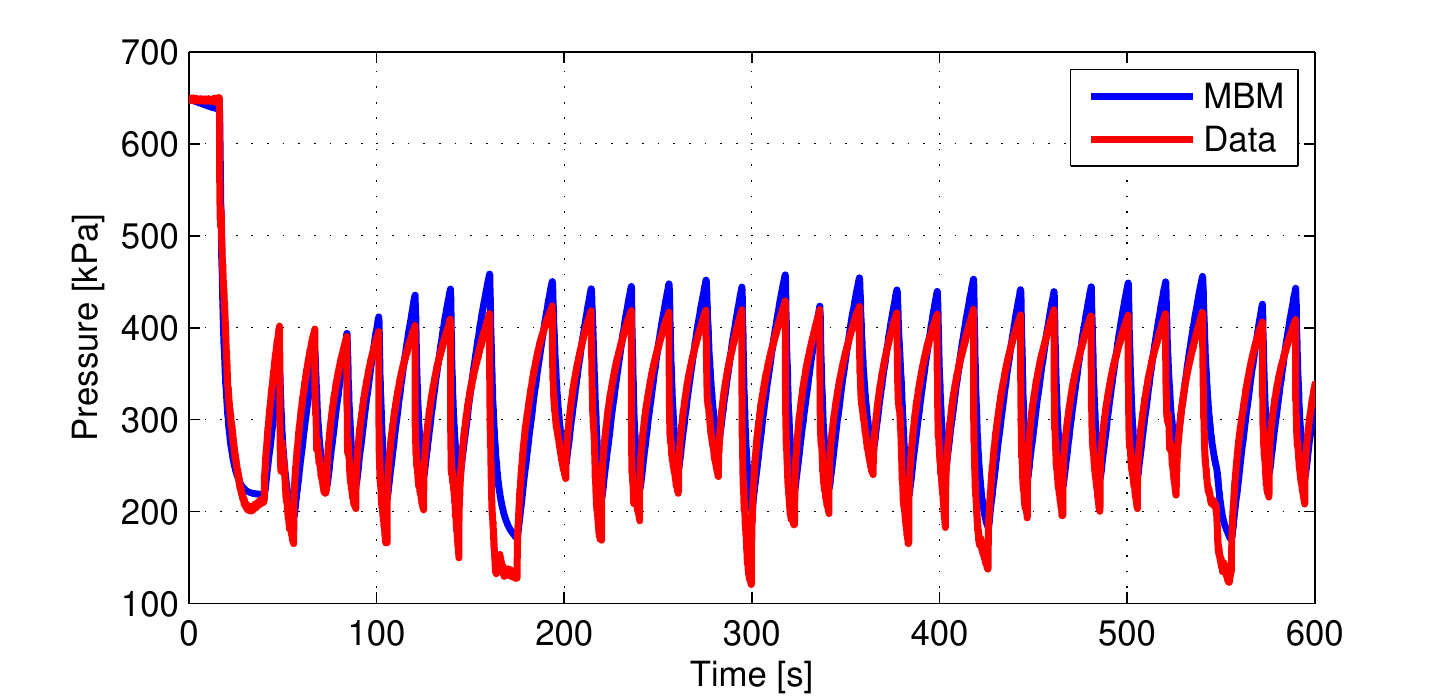}
      \label{fig: Pe_SC03}}
      \caption{Verification of MBM A/C Model for the SC03 driving cycle.}
      \label{fig: Val_SC03}
\end{figure}

The A/C model was calibrated using the data collected during the tests when vehicle/engine speeds are maintained at nominal steady state, and verified  with reference to the SC03 Air Conditioning Cycle in which vehicle speed trace for this regulatory driving cycle is shown in Figure \ref{fig:Veh_SC03}.  The calibration process requires applying multipliers correcting the values of the heat transfer coefficients predicted by the empirical correlations found in the literature. Figure \ref{fig: Val_SC03} also compares the outputs of the model with the corresponding experimental data. During the SC03 test, the compressor speed (related to the engine speed) changes considerably, causing significant variations in the refrigerant flow rate that affect the pressure dynamics in the heat exchangers. This is particularly evident by observing the fluctuations of the condenser pressure, as shown in Figure \ref{fig: Pc_SC03}. From the Root Mean Square Error (RMSE) between calculated pressures and measured pressure, the condenser pressure error is within $6\%$ of its average value, and the evaporator pressure error is around $8\%$.  Therefore, the model appears quite accurate in capturing the pressure dynamics at the condenser and evaporator.

The nonlinear A/C model is linearized at three operating conditions, corresponding to low, medium and high cooling loads, whose controlled inputs and steady state refrigerant status change with respect to the inlet air temperature of the evaporator, as summarized in Table \ref{T:Equilibria}.

\begin{table}[!h]
\small
\caption{A/C Operating Points.}
\label{T:Equilibria}
\centering
\begin{tabular}{|c|c|c|c|c|}
  \hline
  \hline
  $\dot{Q}_a$ & $T_a$ $^0C$ & $N_c$ ($rpm$) & $\alpha$ ($\%$) & $P_e$ ($kpa$)\\ \hline
  Low          & $25$                 & $450$                    & $25$                 & $302.2$ \\ \hline
  Med          & $30$                 & $1000$                   & $40$                 & $251.2$ \\ \hline
  High         & $40$                 & $2500$                   & $55$                 & $204.6$ \\ \hline
  \hline
\end{tabular}

\end{table}

Note that as seen from Figure \ref{fig:GIMC_Scheme}, the coprime factors $\tilde{M}$ and $\tilde{N}$, instead of the plant model $P_{uy}$, are of interest because they are parts of the fault accommodation block. 

\subsection{Coprime factorization of $ H_\infty $ Controller}
$H_{\infty}$ control and estimation problems have been extensively studied in literature \cite{7348666,C6,7355303,J1}, and are considered to be very promising for the automotive industry. In this work, a general $ H_\infty $ synthesis is to find a controller $K$ such that the closed-loop system is asymptotically stable and the $ H_\infty $ norm of the transfer function between the disturbance $\omega$ and controlled output $z$, $ \|T_{\omega z}\|_{\infty} $, is as small as possible \cite{10_zhou1998, 10_zhou1996}.

In order to fit the $H_{\infty}$ synthesis framework, the performance criteria $z$ and unknown disturbances $\omega$ should be clarified for automotive A/C systems. Mathematically, the six elements in the vector of weighted performance criterion $z$ are selected as
\begin{equation}
    z=\begin{bmatrix}
    e_{p_e} & e_{SH} & N_{cmp} & \alpha & p_e & SH
    \end{bmatrix}
    ^T,
\end{equation}
where $e_{p_e} = p_{e,r}- p_e$ and $e_{SH} = SH_r - SH$ are errors on the output evaporator pressure and superheat temperature. $N_{cmp}$ is the compressor rotation speed and $\alpha$ is the valve opening percentage. $p_e$ and $SH$ are, respectively, the evaporator pressure and superheat temperature.

The reference evaporator pressure $p_r$ and superheat temperature $SH_{r}$ are time-varying and regarded as additional disturbances besides the unknown disturbances $\dot m_{ea}$, as well as the noises. Therefore, the disturbance vector $\omega$ is defined as:
\begin{equation}
    w=[\Delta \dot{m}_{ea}, p_{e,r}, SH_r ,n_1, n_2]^T.
\end{equation}

The original A/C model in state-space form is augmented with the output vector and disturbance vector defined, and a $H_{\infty}$ controller is obtained by solving Linear Matrix Inequalities (LMIs) associated with the augmented system according to methods provided in \cite{10_zhou1998, 10_zhou1996}. The above design procedure is detailed in previous work \cite{50_zhang2014JDSMC}, where simulation results are provided to support the validity of the controller during output tracking and disturbance rejection.

Note that as seen from Figure \ref{fig:GIMC_Scheme}, the coprime factors $\tilde{U}$ and $\tilde{V}$, instead of the controller model $K$, are of interest because the fault accommodation signal $q$ is added in between $\tilde{U}$ and $\tilde{V}$. Therefore, the system matrices of the coprime factors are provided below,
\section{Fault Detection and Isolation}
Residuals generated by the FDI module are required to activate the compensator $Q$ in the GIMC structure. The filters for residual generation are designed in the framework of $H_{\infty}$ optimization, and validated using the nonlinear MBM A/C model.
\subsection{$H_\infty$ optimization}
The isolation filter $H_l$ ($l \times p $ transfer matrix) is designed to isolate the fault vector $f$ and decouple the perturbation $d$. The trade-off between these two objectives is also formulated as an optimization problem:
\begin{equation}\label{E:fault isolation}
    \min_{H_I \in RH_{\infty}} \| \left[ \begin{array}{cc} 0 & T \\ \end{array} \right] - H_I \left[ \begin{array}{cc} \tilde{N}_d & \tilde{N}_f \\ \end{array} \right] \|_{\infty},
\end{equation}
where $T \in RH_{\infty}$ is a diagonal transfer matrix to be determined according to the frequency response of $\tilde{N}_f$, in order to achieve the isolation and decoupling objectives.

The above optimization problem can be solved by a Linear Fractional Transformation (LFT):
\begin{equation}\label{E:fault isolation}
    \min_{H_I \in RH_{\infty}} \| F_l(G_{H_I}, H_I) \|_{\infty},
\end{equation}
where $G_{H_I}$ stands for the generalized plant associated to the LFT transformation given by
\begin{equation}
    G_{H_I}(s) =
    \left(
      \begin{array}{ccc}
        0 & T & -I \\
        \tilde{N}_d & \tilde{N}_f & 0 \\
      \end{array}
    \right).
\end{equation}

\subsection{Performance Evaluation}
The fault isolation filter generates independent residuals corresponding to either an actuator fault or a sensor fault. Based on Equation \ref{E:fault isolation}, the performance of a fault isolation filter is determined by the selected weighting function. We next design the fault isolation filter $H_I$ by selecting the following weighting function
\begin{equation}
    T(s) = \left[ \begin{array}{cc} 1 & 0 \\  0 & 1 \end{array}\right]\cdot \frac{1}{10s + 1}.
\end{equation}

\begin{figure}[!htb]
\centering
\includegraphics[width=0.49\textwidth]{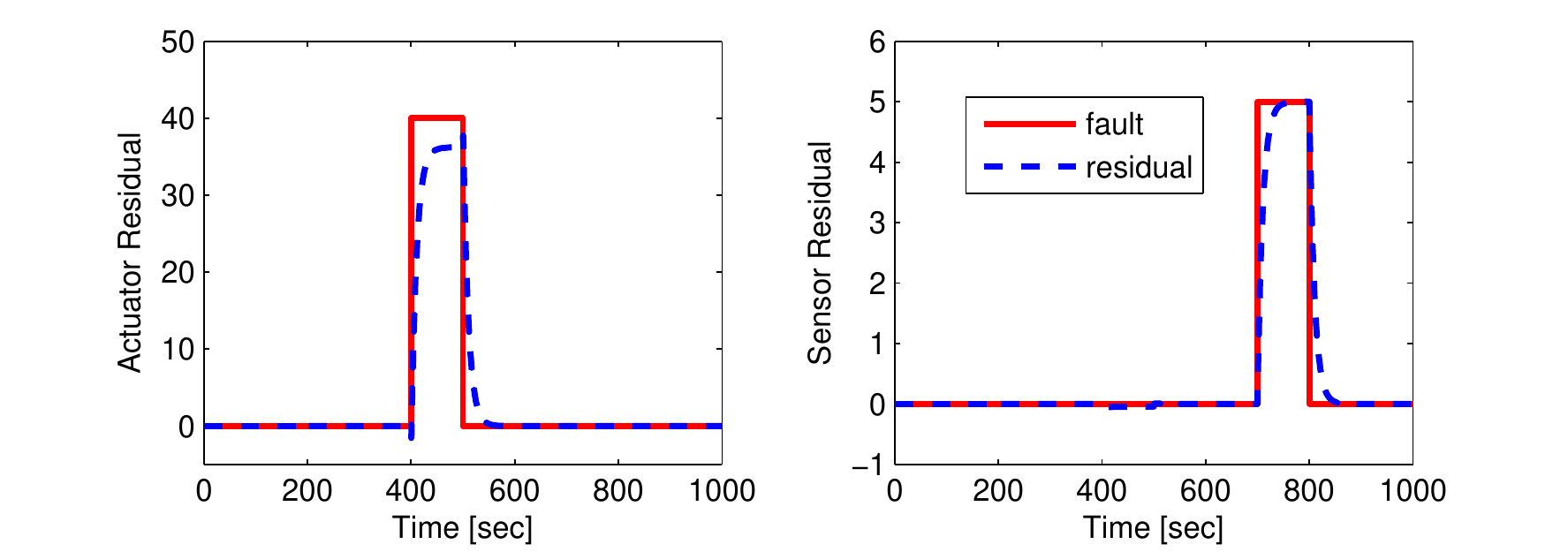}
\caption{Fault Isolation without Disturbance.}
\label{fig:Fault_Isolation_woDisturbance}
\end{figure}

\begin{figure}[!htb]
\centering
\includegraphics[width=0.49\textwidth]{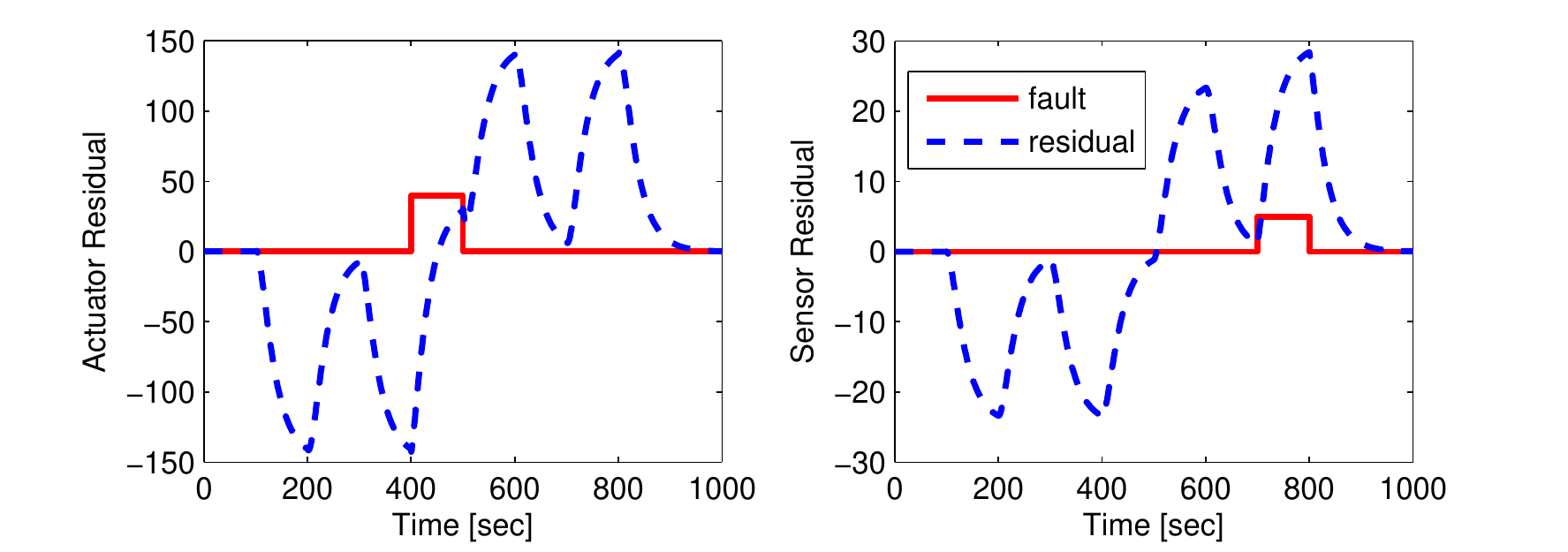}
\caption{Fault Isolation with Disturbance.}
\label{fig:Fault_Isolation_wDisturbance}
\end{figure}

The fault isolation filter $H_I$ calculated by solving an $H_{\infty}$ optimization problem stated in Equation \ref{E:fault isolation} is implemented in the nonlinear closed-loop simulation, where the plant is given by the nonlinear differential equations of the MBM A/C model. This enables the test for model discrepancies with respect to the linear plants used in the filter design process. The actuator fault of $40$ $rpm$ is added into the compressor rotation speed at $400$ second, and the sensor fault of $5 ^oC$ is added into the pressure signal at $700$ second.  When external disturbance is not considered, the two residuals corresponding to actuator fault and sensor fault respectively, are presented in Figure \ref{fig:Fault_Isolation_woDisturbance}. It is clear that both residuals respond to their respective fault, and are decoupled from the other fault. When external disturbance is added as a stepwise signal, however, the two residuals corresponding to actuator fault and sensor fault respectively, as depicted in Figure \ref{fig:Fault_Isolation_wDisturbance}, are very sensitive to the variation of ambient condition on the air side of the heat exchangers.

Sensitivity of the designed isolation filter to external disturbances is due to the fact that the fault observability condition is not met. Current hardware configuration, i.e., one pressure transducer and one thermal couple, is able  to isolate two independent faults simultaneously. In order to ensure the robustness of the isolation filter, it is required to increase the observability of the A/C system. Herein, an additional thermocouple is placed on the wall of the section of the heat exchanger encompassing vapor refrigerant. In other words, an additional state, the temperature of the wall in superheat phase region, is available. Using the same weighting function as before, a new isolation filter $H_I$ is calculated again as an $H_{\infty}$ optimization problem stated in Equation \ref{E:fault isolation}.

\begin{figure}[!htb]
\centering
\includegraphics[width=0.49\textwidth]{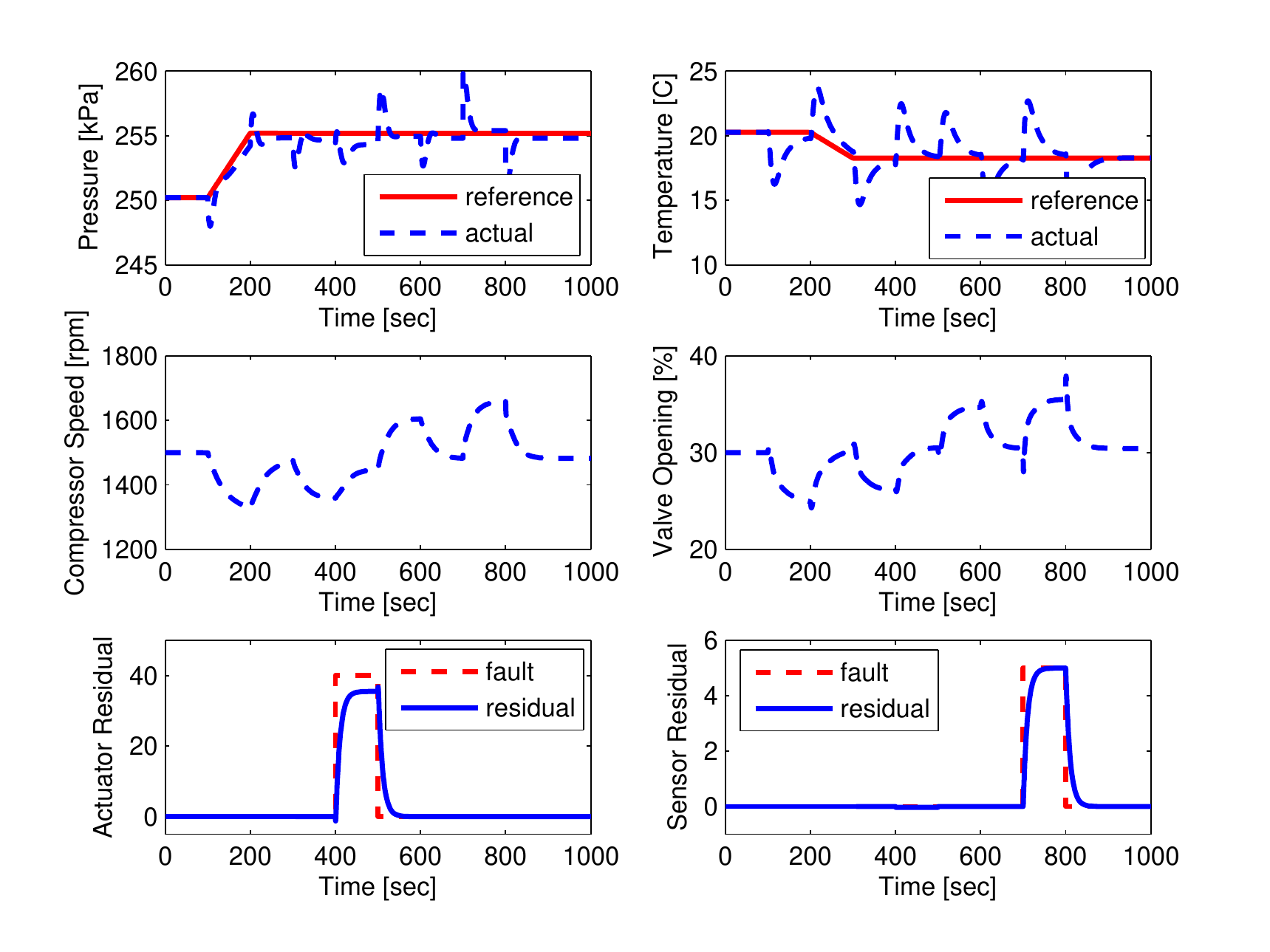}
\caption{Fault Isolation with Disturbance after Additional Sensor Installed.}
\label{fig:Fault_Isolation_Augmented}
\end{figure}

Next, the robustness of the new filter $H_I$ to external disturbance is checked with stepwise variation of boundary conditions on air side of the evaporator as drawn in Figure \ref{fig:Fault_Isolation_Augmented}. A manoeuver is performed that takes the A/C system slightly away from the equilibrium design point. The manoeuver performed takes the A/C system evaporator pressure up $5$ $kpa$ and superheat temperature down $2.5 ^oC$. Due to the existence of external disturbance, the tracking performance is slightly deteriorated as the controller tries to compensate the disturbance during transient. For the same fault setting, however, it is seen that the filter residuals are able to detect and isolate the fault signature with a high degree of accuracy, without significant variation introduced by the external disturbance. Hence, it can be concluded that detection and robustness capability of the filter is guaranteed under current choice of the weight function and computation scheme.

\section{Fault Accommodation}
The GIMC structure is an active FTC scheme since the compensator $Q$ will be activated by the residual signals generated by the FDI module. From analysis of the influence of a fault at different locations of the closed-loop system, the problem of designing a compensator $Q$ is reduced to an optimization problem that minimizes the influence of a fault at either input or output of the plant. The fault accommodation capability of the compensator $Q$ is demonstrated by comparing simulation results of a passive FTC scheme and an active FTC scheme. Moreover, the variation of the compensator $Q$ over the plant operating point is investigated as a preliminary step towards the gain-scheduled GIMC structure.

\subsection{Theoretical Background}
The following lemma originally presented in \cite{10_campos2005} characterizes the dynamic behavior of the control input $u$ and output $y$ of the closed-loop system.

\emph{Lemma 1. } In the GIMC configuration considering additive faults, the resulting closed-loop characteristics for the control signal $u$ and output $y$ are given by
\begin{equation}
\begin{aligned}
     u(s) &= S_i K r(s) - S_i \tilde(V)^{-1} (\tilde{U} \tilde{M}^{-1} + Q)\\
     &\quad\times[\tilde{N}_d d(s) + \tilde{N}_f f(s)], \\
    y(s) &= T_o r(s) + S_o \tilde{M}^{-1} (I-\tilde{N}\tilde{V}^{-1}Q)\\
    &\quad\times[\tilde{N}_d d(s) + \tilde{N}_f f(s)],
\end{aligned}
\end{equation}
where the input sensitivity is $S_i = (I + KP_{uy})^{-1}$, output sensitivity is $S_o = (I + P_{uy}K)^{-1}$ and complementary output sensitivity is $T_o = I - S_o = (I + P_{uy}K)^{-1} P_{uy}K$.


If one desires to attenuate both faults and perturbations at the output $y$, the following optimization scheme is suggested:
\begin{equation}\label{E:fault_actuator}
    \min_{Q \in RH_{\infty}} \| S_o \tilde{M}^{-1} (I - \tilde{N}\tilde{V}^{-1}Q) Q \left[ \begin{array}{cc}  \alpha_d \tilde{N}_d & \alpha_f \tilde{N}_f\\ \end{array} \right] \|_{\infty}.
\end{equation}
The above optimization problem can be solved by a Linear Fractional Transformation (LFT):
\begin{equation}
    \min_{Q \in RH_{\infty}} \| F_l(G_{Q}, Q) \|_{\infty},
\end{equation}
where $G_{Q}$ is given by
\begin{equation}
    G_{Q}(s) =
    \left(
      \begin{array}{ccc}
        \alpha_d S_o P_{dy} & \alpha_f S_o K P_{fy} & - S_o P_{uy} \tilde{V}^{-1} \\
        \alpha_d \tilde{N}_d & \alpha_f \tilde{N}_f & 0 \\
      \end{array}
    \right).
\end{equation}

If one wants to minimize the fault effects on the control signal while reducing the perturbation contribution at the output, the compensator $Q$ should be designed by the following optimization strategy:
\begin{equation}\label{E:fault_sensor}
\begin{aligned}
    &\min_{Q \in RH_{\infty}} \| \left[ \begin{array}{cc} \alpha_d S_o P_{dy} & 0 \\  0 & -\alpha_f S_i K P_{fy} \end{array} \right] \\
    &\quad+ \left[ \begin{array}{c} -\alpha_d S_o P_{uy} \tilde{V}^{-1} \\ -\alpha_f S_i \tilde{V}^{-1}\\ \end{array} \right] Q \left[ \begin{array}{cc}  \tilde{N}_d & \tilde{N}_f\\ \end{array} \right] \|_{\infty}.
    \end{aligned}
\end{equation}
The above optimization problem can be solved by a Linear Fractional Transformation (LFT):
\begin{equation}
    \min_{Q \in RH_{\infty}} \| F_l(G_{Q}, Q) \|_{\infty},
\end{equation}
where $G_{Q}$ represents the generalized plant given by
\begin{equation}
    G_{Q}(s) =
    \left(
      \begin{array}{ccc}
        \alpha_d S_o P_{dy} & 0 & -\alpha_d S_o P_{uy} \tilde{V}^{-1} \\
        0 & -\alpha_f S_i K P_{fy} & -\alpha_f S_i \tilde{V}^{-1} \\
        \tilde{N}_d & \tilde{N}_f & 0 \\
      \end{array}
    \right),
\end{equation}
and $\alpha_d, \alpha_f \in [0,1]$ are two weighting factors to balance  the tradeoff between perturbations and faults reduction.

The two $H_{\infty}$ optimization schemes given in Equations \ref{E:fault_actuator} and \ref{E:fault_sensor} are derived by attenuating the influences of faults on outputs and inputs, respectively. The solutions to the above two problems using the GIMC structure usually generate a compensator $Q$ in high order. Thus it is necessary to perform a controller order reduction. One approach is the standard way, such as balanced truncation appealing to analysis of Hankel norm of every system state. Another approach is to design a specific compensator for every studied fault by replacing the transfer functions $\tilde{N}_f$ and $P_{fy}$ with their corresponding parts.

\subsection{Actuator Fault Accommodation}
When an actuator fault occurs, an active FTC scheme is supposed to ensure the system outputs unchanged, enabling the system inputs to maintain the original values if the steady-state input-output mapping relationships are fixed. Since the sum of the commanded inputs and the faulty input signals are unchanged, the commanded inputs by the controller are modulated automatically to compensate the faulty signals entering the system inputs. Hence, the $H_{\infty}$ optimization schemes given in Equation \ref{E:fault_actuator} are suitable for actuator fault accommodation. After block replacing, the optimization problem for actuator compensator design is defined as:

\begin{equation}
    \min_{Q_{act} \in RH_{\infty}} \| S_o \tilde{M}^{-1} (I - \tilde{N}\tilde{V}^{-1}Q_{act})  \left[ \begin{array}{cc}  \alpha_d \tilde{N}_d & \alpha_f \tilde{N}_f^{act}\\ \end{array} \right] \|_{\infty}.
\end{equation}

\begin{figure}[!h]
\centering
\includegraphics[width=0.49\textwidth]{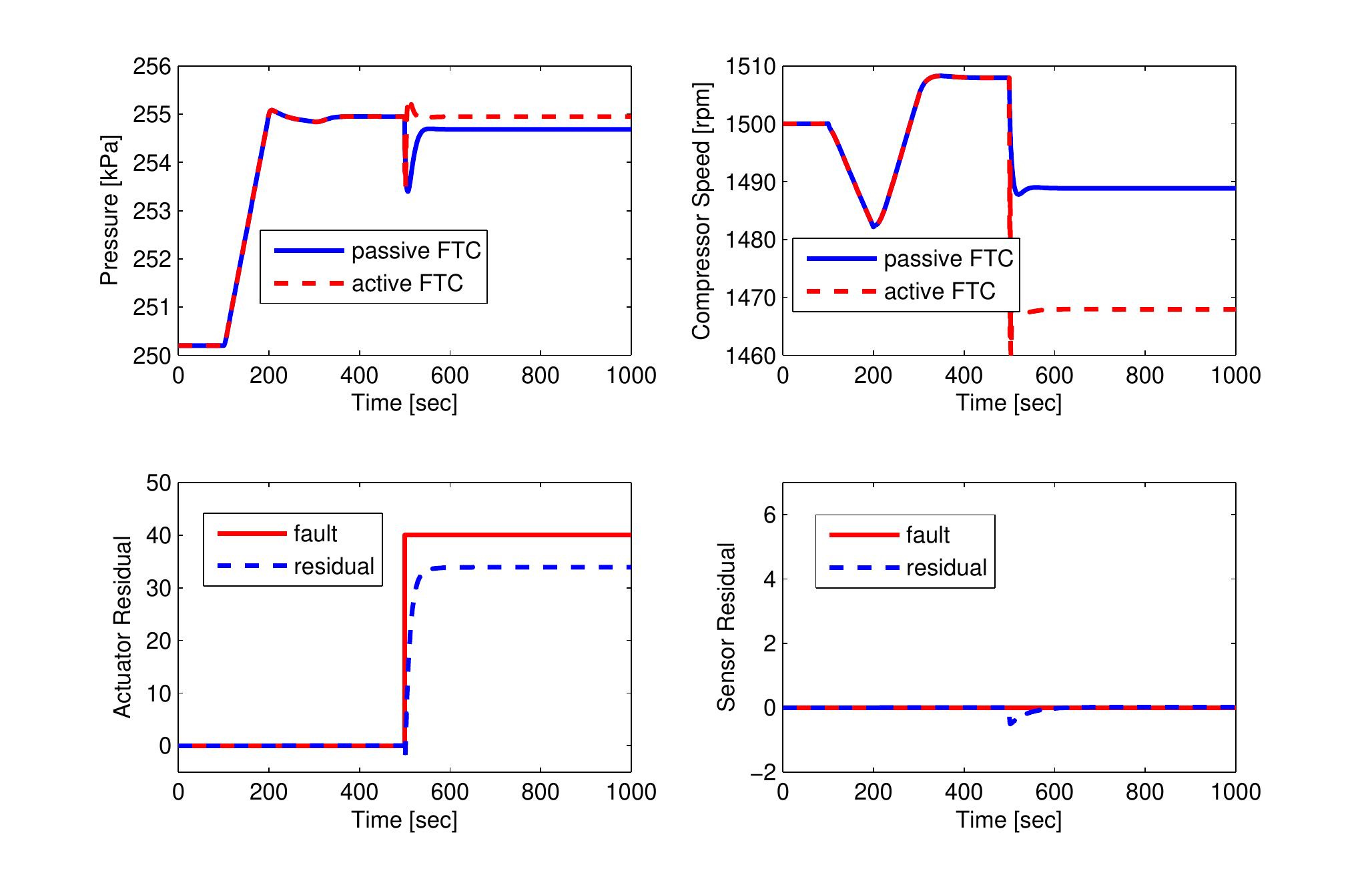}
\caption{Fault Accommodation of Actuator Fault.}
\label{fig:Fault_Accomodation_Actuator}
\end{figure}

The designed compensator $Q_{act}$ is implemented in the FTC scheme shown in Figure \ref{fig:ClosedLoop}, where the plant is replaced with the MBM A/C model.  The simulation results of actuator accommodation are depicted in Figure \ref{fig:Fault_Accomodation_Actuator}, with the same manoeuver strategy as before and one actuator fault of $40$ $rpm$ added at $500$ $rpm$.

The solid lines represent the inputs and the outputs of the A/C system with a passive FTC scheme, which refers to an output tracking controller designed using $H_{\infty}$ synthesis without fault accommodation function. When an actuator fault occurs, a passive FTC scheme has some extent of capability of fault compensation. In the top figures, the compressor speed is reduced by $20$ $rpm$ to ensure that the evaporator pressure does not deviate from the reference value.

The dashed lines represent the inputs and the outputs of the A/C system with an active FTC scheme using GIMC structure. Still from the top figures, the compressor speed is reduced by $40$ $rpm$ by the actuator fault compensator $Q_{act}$, which is the exact amplitude of the actuator fault added. The output evaporator, after minor transient, returns to the reference value without any deviation. The simulation results are natural outcomes of the design process, since the GIMC structure activate a compensator to tolerate the faulty input.

\subsection{Sensor Fault Accommodation}
When a sensor fault occurs, an active FTC scheme is desired to ensure the system inputs unchanged, enabling the system outputs to maintain the original values provided that the steady-state input-output mapping is preserved. Since actual system outputs are unchanged, the sum of the actual outputs and the faulty output signals deviate from the original measured outputs. However, the controller disregards the deviation, and uses the original measured outputs as before. In other words, the sensor fault does not affect the closed-loop system performance significantly. Hence, the $H_{\infty}$ optimization schemes given in Equation \ref{E:fault_sensor} are suitable for actuator fault accommodation. After block replacing, the optimization problem for sensor compensator design is defined as
\begin{equation}
\begin{aligned}
    &\min_{Q_{sen} \in RH_{\infty}} \| \left[ \begin{array}{cc} \alpha_d S_o P_{dy} & 0 \\  0 & -\alpha_f S_i K P_{fy}^{sen} \end{array} \right] \\
    &\quad+ \left[ \begin{array}{c} -\alpha_d S_o P_{uy} \tilde{V}^{-1} \\ -\alpha_f S_i \tilde{V}^{-1}\\ \end{array} \right] Q \left[ \begin{array}{cc}  \tilde{N}_d & \tilde{N}_f^{sen}\\ \end{array} \right] \|_{\infty}.
    \end{aligned}
\end{equation}

\begin{figure}[!h]
\centering
\includegraphics[width=0.49\textwidth]{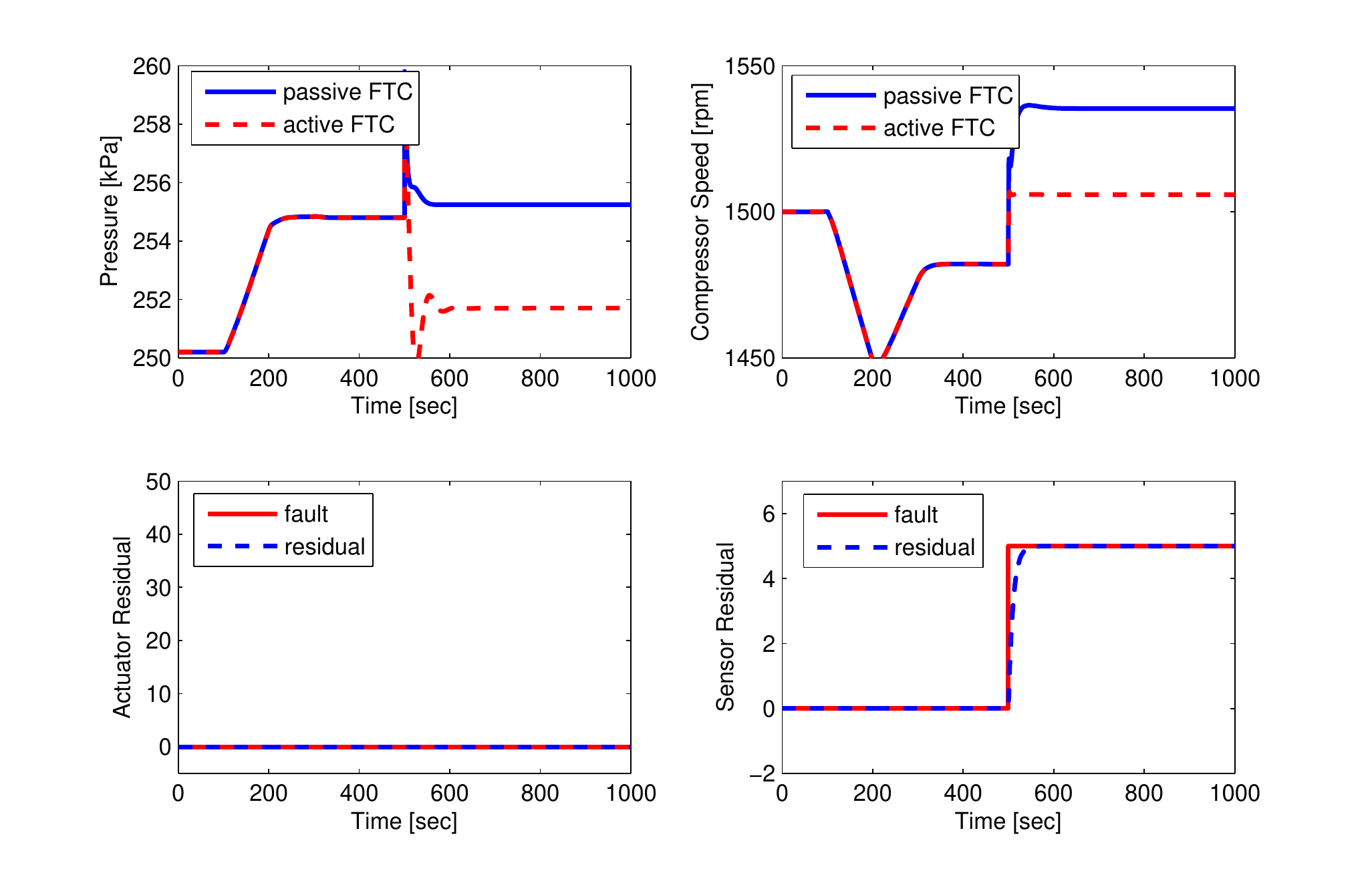}
\caption{Fault Accommodation of Sensor Fault.}
\label{fig:Fault_Accomodation_Sensor}
\end{figure}

The designed compensator $Q_{sen}$ is implemented in the FTC scheme shown in Figure \ref{fig:ClosedLoop}, where the plant is replaced with the MBM A/C model. The simulation results of sensor accommodation are depicted in Figure \ref{fig:Fault_Accomodation_Sensor}, with the same manoeuver strategy as before and one sensor fault of $5$ $kPa$ added at $500$ $sec$.

The solid lines represent the inputs and the outputs of the A/C system with a passive FTC scheme. For a sensor fault, a passive fault tolerant controller uses the faulty measurements to calculate the commanded inputs. In the top figures, the measured evaporator pressure is maintained, while the actual evaporator pressure is enforced to track another reference value deviating from the nominal value by the amplitude of the faulty output, $5 $ $kPa$. Hence, the compressor speed boosts up $50$ $rpm$ in order to drive the actual evaporator pressure to the deviated reference value.

The dashed lines represent the inputs and the outputs of the A/C system with an active FTC scheme using the GIMC structure. Still from the top figures, the compressor speed only boosts up $10$ $rpm$ after the sensor fault compensator $Q_{sen}$ is activate. Since the actual evaporator output is almost unchanged, the measured evaporator pressure starts to deviate from the nominal reference value by $ 4$ $kPa$  after faulty output is added. Although the total sensor fault $5$ $kPa$ is not fully compensated, the active FTC scheme has already achieved much better performance than the passive one. The simulation results are natural outcomes of design process, since the GIMC structure activate a compensator to ensure the actual output is almost unchanged and the measured output changing by the amplitude of the faulty signal.

\subsection{Plant Variation}
\begin{figure}[!h]
\centering
\includegraphics[width=0.49\textwidth]{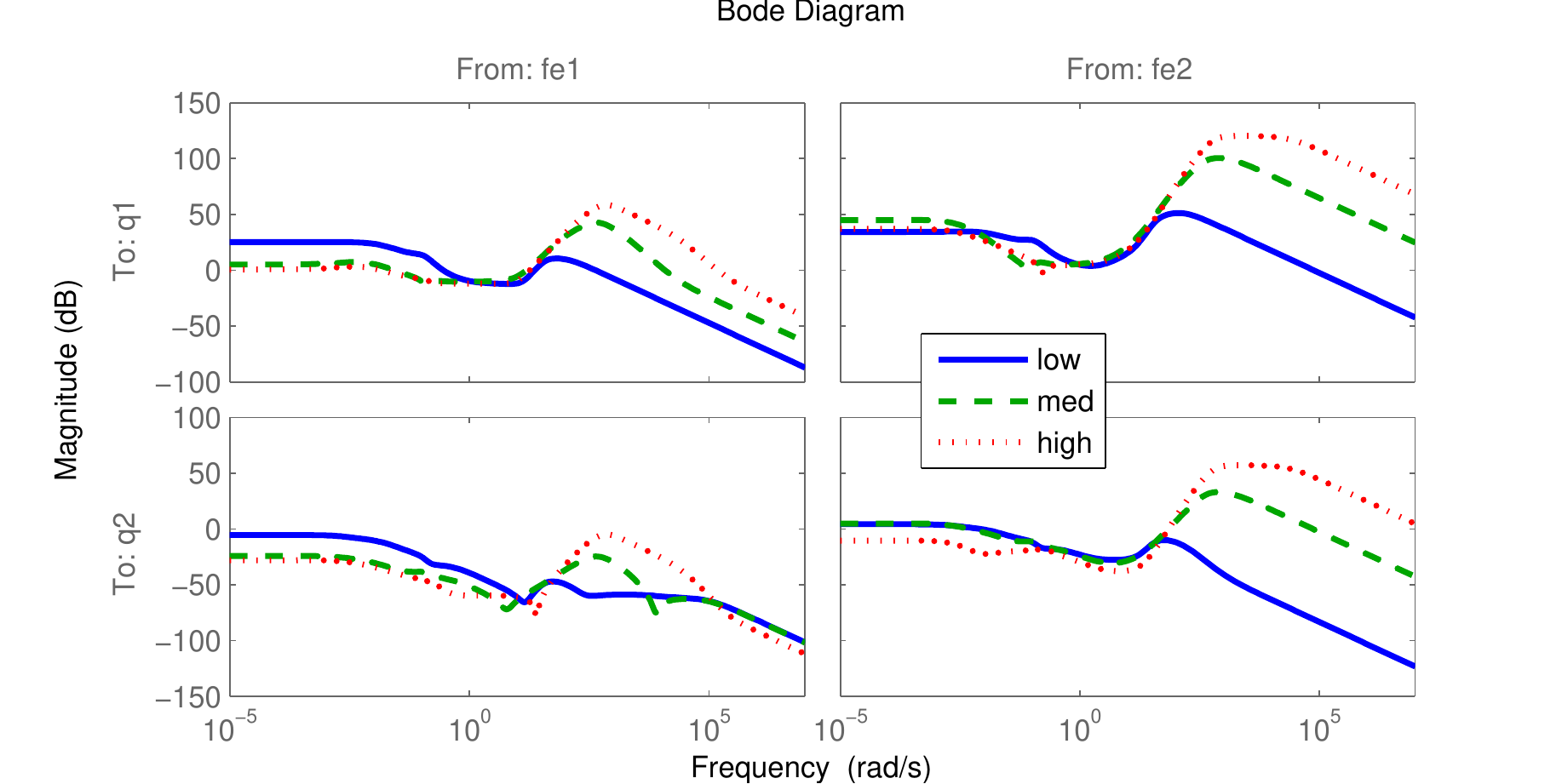}
\caption{Actuator Compensator Variation over Working Points.}
\label{fig:Fault_Accomodation_Qu}
\end{figure}

\begin{figure}[!h]
\centering
\includegraphics[width=0.49\textwidth]{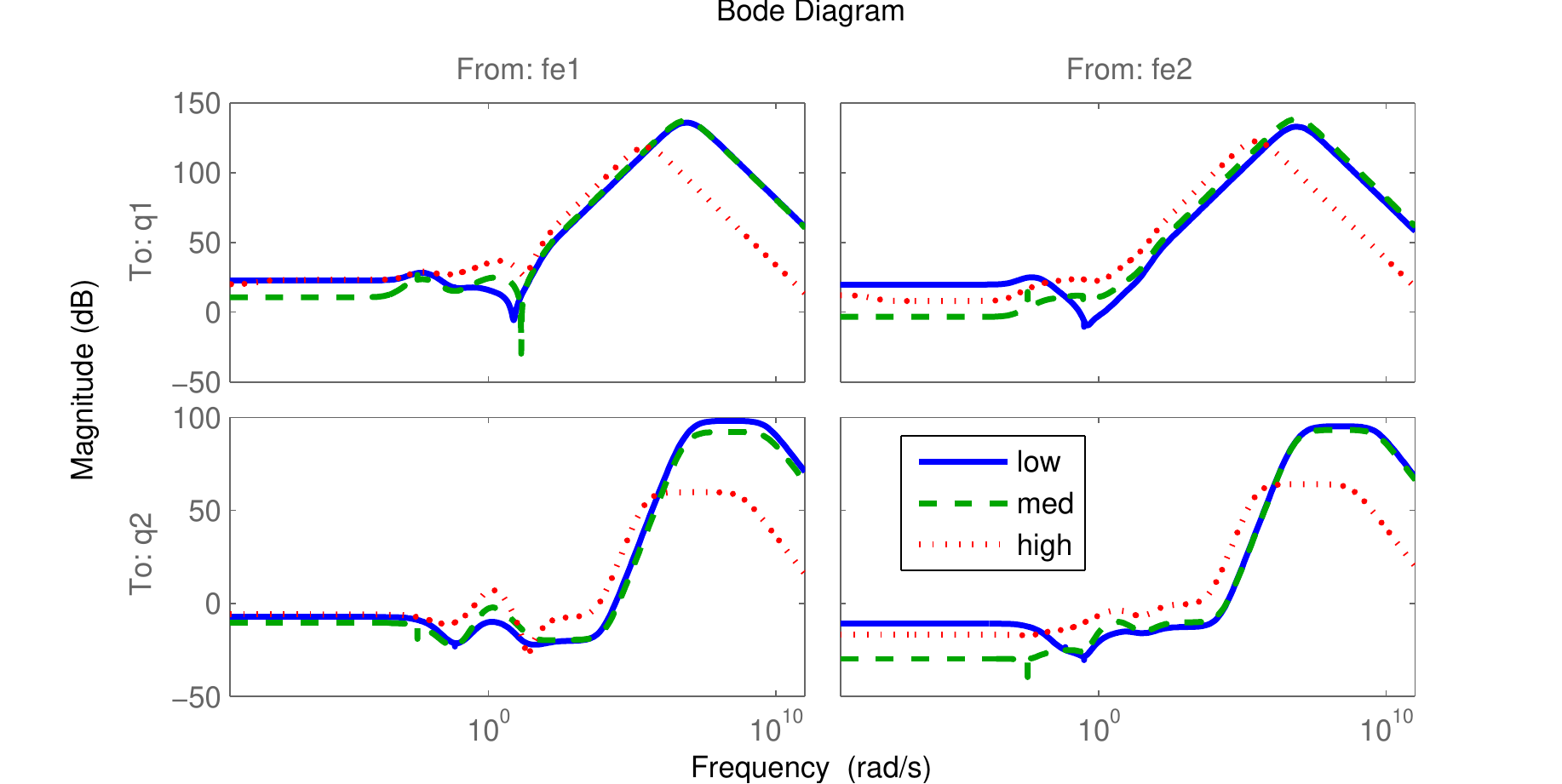}
\caption{Sensor Compensator Variation over Working Points.}
\label{fig:Fault_Accomodation_Qy}
\end{figure}

The variation of the compensator $Q_{act}$ and $Q_{sen}$ over the operating point, or cooling load for A/C system, is examined to exploit the possibility of gain-scheduled GIMC structure. The nonlinear MBM A/C model is linearized at three operating conditions, corresponding to low, medium and high cooling loads. The cooling load is regulated by changing the inlet air temperature of the evaporator. For consistency, the superheat temperature is kept around $20^oC$ by cooperation of the compressor speed $N_c$ and the expansion valve position $\alpha$; however, the evaporator pressure $P_e$ is allowed to vary according to the cooling load as a gain scheduling parameter. Specifically, the boundary conditions, controlled inputs and steady-state refrigeration status are summarized in Table below. For every operating point, the actuator and sensor compensators are designed following the active FTC scheme using the GIMC scheme. From Figures \ref{fig:Fault_Accomodation_Qu} and \ref{fig:Fault_Accomodation_Qy}, it is not difficulty to see that the GIMC can also be made to be adaptive to working point.

\section{Conclusion}

In this paper, the GIMC structure is applied to accommodate actuator and sensor faults of an automotive air conditioning system. The air conditioning system is modeled using the moving boundary method to capture the mixed liquie/vapor flow of the refrigerant in the heat exchangers. The resulting high-order, nonlinear, and control-oriented model is utilized to design an active fault tolerant controller. The designed fault isolation filter is able to isolate actuator and sensor faults when external disturbances are added with an additional thermocouple instrumented on the tube wall. The accommodation performance of the active fault tolerant controller is examined by adding actuator and sensor faults separately. In terms of actuator faults, the FTC scheme using GIMC outperforms passive FTC scheme due to the compensation of the faulty input. As for sensor faults, the deviation of the actual output from the reference output is, thought not completely eliminated, mitigated significantly. The possibility of the FTC using the GIMC scheme by gain-scheduled compensator is also exploited by investigating its variation over operating points. Future work will include the introduction of model uncertainty into the scheme and devise gain-scheduling module.

\section*{Acknowledgment}
This work described in the paper is in part supported by the U.S. Department of Energy, through Chrysler, LLC as the prime contractor. The authors gratefully acknowledge Chrysler, LLC and Dr. Timothy C. Scott for providing the data to calibrate the model and for the useful discussions.

%
%
%
%
%

\bibliographystyle{ieeetr}
\bibliography{ACbiblio}

\begin{thebibliography}{10}

\bibitem{50_chen2012}
J.~Chen and R.~J. Patton, {\em Robust model-based fault diagnosis for dynamic
  systems}.
\newblock Springer Publishing Company, Incorporated, 2012.

\bibitem{50_isermann1997}
R.~Isermann and P.~Balle, ``Trends in the application of model-based fault
  detection and diagnosis of technical processes,'' {\em Control engineering
  practice}, vol.~5, no.~5, pp.~709--719, 1997.

\bibitem{50_isermann2006}
R.~Isermann, {\em Fault-diagnosis systems}.
\newblock Springer, 2006.

\bibitem{50_blanke2006}
M.~Blanke and J.~Schr{\"o}der, {\em Diagnosis and fault-tolerant control},
  vol.~2.
\newblock Springer, 2006.

\bibitem{10_zhang2008}
Y.~Zhang and J.~Jiang, ``Bibliographical review on reconfigurable
  fault-tolerant control systems,'' {\em Annual reviews in control}, vol.~32,
  no.~2, pp.~229--252, 2008.

\bibitem{J2}
Z.~Li, I.~V. Kolmanovsky, U.~V. Kalabic, E.~M. Atkins, J.~Lu, and D.~P. Filev,
  ``Optimal state estimation for systems driven by jump-diffusion process with
  application to road anomaly detection,'' {\em IEEE Transactions on Control
  Systems Technology}, 2016.
\newblock doi:10.1109/TCST.2016.2620062.

\bibitem{J3}
Z.~Li, I.~Kolmanovsky, E.~Atkins, J.~Lu, D.~P. Filev, and J.~Michelini, ``Road
  risk modeling and cloud-aided safety-based route planning,'' {\em IEEE
  Transactions on Cybernetics}, vol.~46, pp.~2473--2483, Nov 2016.

\bibitem{6214702}
P.~B. Sujit, D.~E. Lucani, and J.~B. Sousa, ``Bridging cooperative sensing and
  route planning of autonomous vehicles,'' {\em IEEE Journal on Selected Areas
  in Communications}, vol.~30, pp.~912--922, June 2012.

\bibitem{J8}
Z.~Li, D.~P. Filev, I.~Kolmanovsky, E.~Atkins, and J.~Lu, ``A new clustering
  algorithm for processing gps-based road anomaly reports with a mahalanobis
  distance,'' {\em IEEE Transactions on Intelligent Transportation Systems},
  2016.
\newblock doi:10.1109/TITS.2016.2614350.

\bibitem{C7}
X.~Yin, Z.~Li, S.~L. Shah, L.~Zhang, and C.~Wang, ``Fuel efficiency modeling
  and prediction for automotive vehicles: A data-driven approach,'' in {\em
  2015 IEEE International Conference on Systems, Man, and Cybernetics},
  pp.~2527--2532, Oct 2015.

\bibitem{C1}
Z.~Li, I.~Kolmanovsky, E.~Atkins, J.~Lu, and D.~Filev, ``Road anomaly
  estimation: Model based pothole detection,'' in {\em 2015 American Control
  Conference (ACC)}, pp.~1315--1320, July 2015.

\bibitem{C2}
Z.~Li, I.~Kolmanovsky, E.~Atkins, J.~Lu, D.~Filev, and J.~Michelini, ``Cloud
  aided safety-based route planning,'' in {\em 2014 IEEE International
  Conference on Systems, Man, and Cybernetics}, pp.~2495--2500, Oct 2014.

\bibitem{1219456}
L.~H. Lee, K.~C. Tan, K.~Ou, and Y.~H. Chew, ``Vehicle capacity planning
  system: a case study on vehicle routing problem with time windows,'' {\em
  IEEE Transactions on Systems, Man, and Cybernetics - Part A: Systems and
  Humans}, vol.~33, pp.~169--178, March 2003.

\bibitem{C3}
Z.~Li, I.~Kolmanovsky, E.~Atkins, J.~Lu, D.~Filev, and J.~Michelini, ``Cloud
  aided semi-active suspension control,'' in {\em Computational Intelligence in
  Vehicles and Transportation Systems (CIVTS), 2014 IEEE Symposium on},
  pp.~76--83, Dec 2014.

\bibitem{J7}
X.~Yin, L.~Zhang, Y.~Zhu, C.~Wang, and Z.~Li, ``Robust control of networked
  systems with variable communication capabilities and application to a
  semi-active suspension system,'' {\em IEEE/ASME Transactions on
  Mechatronics}, vol.~21, pp.~2097--2107, Aug 2016.
\newblock doi:10.1109/TMECH.2016.2553522.

\bibitem{7506101}
B.~Liu, M.~Saif, and H.~Fan, ``Adaptive fault tolerant control of a half-car
  active suspension systems subject to random actuator failures,'' {\em
  IEEE/ASME Transactions on Mechatronics}, vol.~21, pp.~2847--2857, Dec 2016.

\bibitem{C9}
Z.~Li, H.~R. Ossareh, I.~V. Kolmanovsky, E.~M. Atkins, and J.~Lu, ``Nonlinear
  control of semi-active suspension systems: A quasi-linear control approach,''
  in {\em 2016 American Control Conference (ACC)}, pp.~2397--2402, July 2016.

\bibitem{7533442}
V.~S. Deshpande, P.~D. Shendge, and S.~B. Phadke, ``Nonlinear control for dual
  objective active suspension systems,'' {\em IEEE Transactions on Intelligent
  Transportation Systems}, 2016.
\newblock doi:10.1109/TITS.2016.2585343.

\bibitem{J9}
X.~Yin, L.~Zhang, Z.~Ning, D.~Tian, A.~Alsaedi, and B.~Ahmad, ``State
  estimation via {M}arkov switching-channel network and application to
  suspension systems,'' {\em IET Control Theory \& Applications}, 2016.
\newblock doi:10.1049/iet-cta.2016.1108.

\bibitem{C4}
Z.~Li, I.~Kolmanovsky, E.~Atkins, J.~Lu, and D.~Filev, ``${H}_{\infty}$
  filtering for cloud-aided semi-active suspension with delayed road
  information,'' {\em IFAC-PapersOnLine}, vol.~48, no.~12, pp.~275--280, 2015.
\newblock 12th \{IFAC\} Workshop on Time Delay Systems, Ann Arbor, Michigan.

\bibitem{10_katipamula2005p1}
S.~Katipamula and M.~R. Brambley, ``Review article: methods for fault
  detection, diagnostics, and prognostics for building systems—a review, part
  i,'' {\em HVAC\&R Research}, vol.~11, no.~1, pp.~3--25, 2005.

\bibitem{10_keir2006}
M.~C. Keir and A.~G. Alleyne, ``Dynamic modeling, control, and fault detection
  in vapor compression systems,'' tech. rep., Air Conditioning and
  Refrigeration Center. College of Engineering. University of Illinois at
  Urbana-Champaign., 2006.

\bibitem{10_janecke2011}
A.~K. Janecke, {\em A Comparison of Fault Detection Methods For a Transcritical
  Refrigeration System}.
\newblock PhD thesis, Texas A\&M University, 2011.

\bibitem{10_wagner1992}
J.~Wagner and R.~Shoureshi, ``Failure detection diagnostics for thermofluid
  systems,'' {\em Journal of dynamic systems, measurement, and control},
  vol.~114, no.~4, pp.~699--706, 1992.

\bibitem{10_lee1996}
W.-Y. Lee, C.~Park, and G.~E. Kelly, ``Fault detection in an air-handling unit
  using residual and recursive parameter identification methods,'' {\em
  Transactions-American Society Of Heating Refrigerating And Air Conditioning
  Engineers}, vol.~102, pp.~528--539, 1996.

\bibitem{01_he1997}
X.-D. He, S.~Liu, and H.~H. Asada, ``Modeling of vapor compression cycles for
  multivariable feedback control of hvac systems,'' {\em Journal of dynamic
  systems, measurement, and control}, vol.~119, no.~2, pp.~183--191, 1997.

\bibitem{01_Asada1998}
X.~He, S.~Liu, H.~Asada, and H.~Itoh, ``Multivariable control of vapor
  compression systems,'' {\em HVAC\&R Research}, vol.~4, no.~3, pp.~205--230,
  1998.

\bibitem{02_Li2010}
B.~Li and A.~Alleyne, ``A dynamic model of a vapor compression cycle with
  shut-down and start-up operations,'' {\em International Journal of
  refrigeration}, vol.~33, no.~3, pp.~538--552, 2010.

\bibitem{10_nett1988}
C.~Nett, C.~Jacobson, and A.~Miller, ``An integrated approach to controls and
  diagnostics: The 4-parameter controller,'' in {\em American Control
  Conference, 1988}, pp.~824--835, IEEE, 1988.

\bibitem{10_niemann1997}
H.~Niemann and J.~Stoustrup, ``Integration of control and fault detection:
  nominal and robust design,'' {\em IFAC Fault Detection, Supervision and
  Safety for Technical Processes. Hull, UK}, pp.~341--346, 1997.

\bibitem{10_stoustrup1997}
J.~Stoustrup, M.~Grimble, and H.~Niemann, ``Design of integrated systems for
  the control and detection of actuator/sensor faults,'' {\em Sensor Review},
  vol.~17, no.~2, pp.~138--149, 1997.

\bibitem{10_marcos2005}
A.~Marcos and G.~J. Balas, ``A robust integrated controller/diagnosis aircraft
  application,'' {\em International Journal of Robust and Nonlinear Control},
  vol.~15, no.~12, pp.~531--551, 2005.

\bibitem{10_zhou2001}
K.~Zhou and Z.~Ren, ``A new controller architecture for high performance,
  robust, and fault-tolerant control,'' {\em Automatic Control, IEEE
  Transactions on}, vol.~46, no.~10, pp.~1613--1618, 2001.

\bibitem{10_campos2003}
D.~U. Campos-Delgado and K.~Zhou, ``Reconfigurable fault-tolerant control using
  gimc structure,'' {\em Automatic Control, IEEE Transactions on}, vol.~48,
  no.~5, pp.~832--839, 2003.

\bibitem{10_campos2005}
D.~U. Campos-Delgado, S.~Martinez-Martinez, and K.~Zhou, ``Integrated
  fault-tolerant scheme for a dc speed drive,'' {\em Mechatronics, IEEE/ASME
  Transactions on}, vol.~10, no.~4, pp.~419--427, 2005.

\bibitem{10_campos2008}
D.~Campos-Delgado, E.~Palacios, and D.~Espinoza-Trejo, ``Fault detection,
  isolation, and accommodation for lti systems based on gimc structure,'' {\em
  Journal of Control Science and Engineering}, vol.~2008, p.~9, 2008.

\bibitem{10_zhang2014}
Q.~Zhang, L.~Fiorentini, and M.~Canova, ``H∞ robust control of an automotive
  air conditioning system,'' in {\em American Control Conference (ACC), 2014},
  pp.~5675--5680, IEEE, 2014.

\bibitem{50_zhang2014JDSMC}
Q.~Zhang and M.~Canova, ``Output feedback control of automotive air
  conditioning system using $h_{\infty}$ technique,'' {\em Accepted by
  International Journal of Refrigeration}, 2015.

\bibitem{7348666}
Y.~Tang, H.~Gao, and J.~Kurths, ``Robust ${H}_{\infty }$ self-triggered control
  of networked systems under packet dropouts,'' {\em IEEE Transactions on
  Cybernetics}, vol.~46, pp.~3294--3305, Dec 2016.

\bibitem{C6}
Z.~Li, X.~Yin, X.~Yin, Y.~Xie, and C.~Wang, ``Distributed ${H}_{\infty}$
  filtering over multiple-channel sensor networks with markovian channel
  switching and time-varying delays,'' in {\em 2015 54th IEEE Conference on
  Decision and Control (CDC)}, pp.~7410--7415, Dec 2015.

\bibitem{7355303}
A.~Selivanov and E.~Fridman, ``Event-triggered ${H}_{\infty }$ control: A
  switching approach,'' {\em IEEE Transactions on Automatic Control}, vol.~61,
  pp.~3221--3226, Oct 2016.

\bibitem{J1}
X.~Yin, Z.~Li, L.~Zhang, and M.~Han, ``Distributed state estimation of
  sensor-network systems subject to markovian channel switching with
  application to a chemical process,'' {\em IEEE Transactions on Systems, Man,
  and Cybernetics: Systems}, 2016.
\newblock doi:10.1109/TSMC.2016.2632155.

\bibitem{10_zhou1998}
K.~Zhou and J.~C. Doyle, {\em Essentials of robust control}, vol.~104.
\newblock Prentice hall Upper Saddle River, NJ, 1998.

\bibitem{10_zhou1996}
K.~Zhou, J.~C. Doyle, K.~Glover, {\em et~al.}, {\em Robust and optimal
  control}, vol.~40.
\newblock Prentice Hall New Jersey, 1996.

\end{thebibliography}
\end{document}